\documentclass[12pt,letter]{article}
\pdfoutput=1
\usepackage{graphicx, epsfig, color,cite}
\usepackage{amsmath}
\usepackage{amssymb}
\usepackage{float}
\usepackage{caption,subcaption,graphicx}
\usepackage{hyperref}

\newcommand\snowmass{\begin{center}\rule[-0.2in]{\hsize}{0.01in}\\\rule{\hsize}{0.01in}\\
\vskip 0.1in Submitted to the  Proceedings of the US Community Study\\ 
on the Future of Particle Physics (Snowmass 2021)\\ 
\rule{\hsize}{0.01in}\\\rule[+0.2in]{\hsize}{0.01in} \end{center}}

\def\to{\rightarrow}

\def\bi{\begin{itemize}}
\def\ei{\end{itemize}}

\def\tchi{\tilde\chi}

\def\tu{\tilde u}

\def\tst{\tilde t}

\def\tg{\tilde g}

\def\tq{\tilde q}
\def\alt{\lesssim}
\def\agt{\gtrsim}
\def\be{\begin{equation}}  
\def\ee{\end{equation}}  
\def\bea{\begin{eqnarray}}  
\def\eea{\end{eqnarray}}

\begin{document}
\begin{titlepage}
\begin{flushright}
OU-HEP-220301
\end{flushright}

\vspace{0.5cm}
\begin{center}
  {\Large \bf Mini-review: Expectations for supersymmetry\\
    from the string landscape}\\
  \snowmass
\vspace{1.2cm} \renewcommand{\thefootnote}{\fnsymbol{footnote}}
{\large Howard Baer$^{1}$\footnote[1]{Email: baer@ou.edu },
Vernon Barger$^2$\footnote[2]{Email: barger@pheno.wisc.edu},
Shadman Salam$^1$\footnote[3]{Email: shadman.salam@ou.edu} and
Dibyashree Sengupta$^3$\footnote[4]{Email: dsengupta@phys.ntu.edu.tw}
}\\ 
%and Dibyashree Sengupta$^1$\footnote[4]{Email: Dibyashree.Sengupta-1@ou.edu}
\vspace{1.2cm} \renewcommand{\thefootnote}{\arabic{footnote}}
{\it 
$^1$Homer L. Dodge Department of Physics and Astronomy,
University of Oklahoma, Norman, OK 73019, USA \\[3pt]
}
{\it 
$^2$Department of Physics,
University of Wisconsin, Madison, WI 53706 USA \\[3pt]
}
{\it 
$^3$Department of Physics,
National Taiwan University, Taipei, Taiwan 10617, R.O.C. \\[3pt]
}

\end{center}

\vspace{0.5cm}
\begin{abstract}
\noindent
In this mini-review, we summarize a variety of findings pertaining to
consequences of the landscape of string theory for supersymmetry (SUSY)
phenomenology. The idea is to adopt the MSSM as the most parsimonious
$4-d$ low energy EFT after string compactification but where the scale of
SUSY breaking is as yet undetermined.
A power-law landscape draw to large soft terms is tempered by the
requirement that the derived value of the weak scale lie within the
anthropic window of Agrawal {\it et al.} (ABDS).
Such a set-up predicts a light Higgs mass
$m_h\sim 125$ GeV with sparticles generally beyond LHC bounds.
We discuss consequences for LHC searches: light higgsinos, highly mixed
TeV-scale top squarks, same-sign diboson events and $m_{\tg}\sim 2-5$ TeV.
We expect dark matter to consist of an axion/higgsino-like WIMP admixture.
\end{abstract}
\end{titlepage}
%\pacs{12.60.-i, 95.35.+d, 14.80.Ly, 11.30.Pb}
%12.60.-i   Models beyond the standard model
%95.35.+d   Dark matter

\section{Introduction and set up}
\label{sec:intro}

So far, superstring theory provides our only successful unification of
quantum mechanics with general relativity\cite{Green:1987sp,Green:1987mn}.
However, to avoid anomalies,
then the dimensionality of spacetime must be increased to ten (or even eleven in
$M$-theory\cite{Horava:1996ma}).
To gain accord with the $4-d$ physics of the Standard Model,
one must compactify the extra six dimensions on a suitable compact
manifold, such a Calabi-Yau (CY) manifold, which preserves $N=1$
supersymmetry in the low energy effective field theory (EFT)\cite{Candelas:1985en}.
The $4-d$ laws of physics after compactification then depend on various
properties of the compact manifold, which are parametrized by
vacuum expectation values (vevs) of
moduli fields: gravitationally interacting scalar fields with a flat
classical potential. To gain a realistic and potentially predictive
theory, the moduli must be stabilized so as to avoid runaway solutions
and to gain well-determined vevs which in turn determine various
low energy properties such as the values of gauge and Yukawa
couplings and soft SUSY breaking terms\cite{Conlon:2015uua}.

At the turn of the 21st century, it was realized that the number of
possibilities for compact CY manifolds was far, far greater than
anticipated\cite{Bousso:2000xa}.
Under flux comactifications\cite{Douglas:2006es},
the number of distinct vacuum states might range
from $10^{500}$\cite{Ashok:2003gk} to $10^{272,000}$\cite{Taylor:2015xtz}.
These are more than enough to implement Weinberg's anthropic solution to
the cosmological constant (CC) problem\cite{Weinberg:1987dv}.
It also provides a new understanding of how vastly different energy scales
may emerge from string theory which contains only the string scale $m_s$
at its most fundamental level.
In an eternally inflating multiverse\cite{Guth:2000ka,Linde:2015edk},
different pocket
universes can arise with different physical constants.
Requiring a pocket universe to allow for large scale structure in the
form of galaxies and clusters, then only those with a tiny cosmological
constant are allowed. Indeed, this approach allowed Weinberg to predict the
value of the cosmological constant to a factor of a few a decade before
its tiny yet non-zero value was measured\cite{Weinberg:1987dv,Martel:1997vi}.

A similar approach may be applied to the origin of the weak scale
$m_{weak}\sim m_{W,Z,h}$ in models with weak scale supersymmetry\cite{Baer:2006rs} (WSS) (for a recent review of WSS after LHC Run 2, see Ref. \cite{Baer:2020kwz}).
In models such as the
MSSM, the weak scale is determined by the soft SUSY breaking parameters
and the SUSY conserving $\mu$ parameter\footnote{Twenty solutions to the SUSY $\mu$ problem are reviewed in Ref. \cite{Bae:2019dgg}.}. Minimizing the MSSM scalar (Higgs)
potential, one finds
\be
m_Z^2/2=\frac{m_{H_d}^2+\Sigma_d^d-(m_{H_u}^2+\Sigma_u^u)\tan^2\beta}{\tan^2\beta -1}-\mu^2
\label{eq:mzs}
\ee
Here, $m_{H_u}^2$ and $m_{H_d}^2$ are the {\it weak scale values} of the
soft SUSY breaking Higgs masses, $\tan\beta \equiv v_u/v_d$ is the ratio of
Higgs field vevs and the terms $\Sigma_u^u$ and $\Sigma_d^d$ contain over
40 loop corrections to the Higgs potential (for a tabulation, see
Ref's \cite{Baer:2012cf} and \cite{Baer:2021tta}).
The largest of these typically come from the top squarks:
$\Sigma_u^u(\tst_{1,2})$. Equation~\ref{eq:mzs} allows for a definition of
the weak scale finetuning measure $\Delta_{EW}\equiv |largest\ term\ on\ RHS\ |
/(m_Z^2/2)$\cite{Baer:2012up}.

If we adopt a so-called {\it fertile patch} of the landscape-- those vacua
whose $4-d$ low energy EFT is, by parsimony, the MSSM--
then we would expect the magnitude of the weak scale to vary in each
pocket universe depending on the values of the soft breaking terms
and $\mu$ parameter in that same pocket
universe\cite{Baer:2020vad}.
We can write the distribution of vacua versus the soft SUSY breaking
scale $m_{soft}$ (where $m_{soft}\simeq m_{SUSY}^2/m_P$ in gravity mediation where
$m_{SUSY}\sim 10^{11}$ GeV is the mass scale associated with hidden sector SUSY
breaking) as
\be
dN_{vac}\sim f_{SUSY}(m_{soft})\cdot f_{EWSB}(m_{soft})\cdot dm_{soft} .
\label{eq:dNvac}
\ee
Douglas and others\cite{Douglas:2004qg,Susskind:2004uv,Arkani-Hamed:2005zuc}
originally proposed that
\be
f_{SUSY}\sim m_{soft}^{2n_F+n_D-1}
\ee
where $n_F$ is the number of $F$-term SUSY
breaking fields and $n_D$ is the number of $D$-term SUSY breaking fields
contributing to the overall SUSY breaking scale.
The factor 2 arises since it is expected that the vevs $F_i$ are distributed
uniformly as complex numbers whilst the $D_j$ are distributed
uniformly as real numbers. It was realized shortly thereafter that this
may be too simplistic in that the sources of SUSY breaking may not all be
independent\cite{Denef:2004cf}.
But even so, for the textbook case of spontaneous SUSY
breaking by a single $F$-term, then already one expects soft terms to be
statistically favored to large values by a linear distribution
$f_{SUSY}\sim m_{soft}^1$.
An alternative-- emphasized by Dine {\it et al.}\cite{Dine:2004is,Dine:2005iw}-- is that
SUSY is broken non-perturbatively in a hidden sector via {\it e.g.}
dynamical SUSY breaking either via gaugino condensation or via instanton
effects. In such a case, then no SUSY breaking scale is favored over any other,
which would result in $f_{SUSY}\sim m_{soft}^{-1}$.
These results have some further support from Broeckel {\it et al.}\cite{Broeckel:2020fdz}
where they investigate the statistics of SUSY breaking in the landscape
including considerations of K\"ahler moduli ($T_i$) stabilization. For the
large volume scenario (LVS)\cite{Balasubramanian:2005zx} where the $T_i$ are stabilized by a balance between
perturbative and non-perturbative effects, then they find
$f_{SUSY}\sim m_{soft}^{-1}$ while for non-perturbative K\"ahler moduli
stabilization as in KKLT\cite{Kachru:2003aw} they find $f_{SUSY}\sim m_{soft}^1$.
Thus, in the following, we will compare statistical predictions from the
string landscape assuming $f_{SUSY}\sim m_{soft}^{\pm 1}$.

The other relevant distribution contains (anthropic) selection effects
in $f_{EWSB}$. Agrawal, Barr, Donoghue and Seckel (ABDS)\cite{Agrawal:1997gf,Agrawal:1998xa}
showed already in 1997 that too large a
value of the weak scale in various causally disconnected domains of the
multiverse would lead, via the up-down quark mass difference, to unstable
nuclei and lack of atoms which are apparently needed for life as we know it.
%(The existence of atoms for life as we know it is known as the {\it atomic prin%ciple}.) 
They estimate that for pocket universes with
$m_{weak}^{PU}\agt (2-5) m_{weak}^{OU}$, then one is in violation of this so-called {\it atomic principle} (where $m_{weak}^{OU}$ corresponds to the magnitude of the weak scale in our universe). Without finetuning, then the pocket universe
value of the weak scale corresponds to the maximal term on the RHS of
Eq. \ref{eq:mzs}. Then requiring for definiteness $m_{weak}^{PU}<4m_{weak}^{OU}$
corresponds to $\Delta_{EW}\alt 30$. Allowing for finetuning (where
$m_Z$ is not hardwired to 91.2 GeV) then much the same results are
obtained in Ref.~\cite{Baer:2022wxe}.

\section{Sparticle and Higgs mass distributions from the landscape: $n=\pm 1$}
\label{sec:plots}

In our present discussion, we will adopt the 3 extra parameter
non-universal Higgs model (NUHM3) for explicit calculations\cite{Ellis:2002wv,Ellis:2002iu,Baer:2005bu}.
In this model, the matter scalars of the first two generations
are assumed to live in the 16-dimensional spinor of $SO(10)$
as is expected in string models exhibiting {\it local}
grand unification, where different gauge groups are present at different
locales on the compactified manifold\cite{Buchmuller:2005sh}.
In this case, it is really expected that each generation acquires different
$m_0(1)$, $m_0(2)$ and $m_0(3)$ soft breaking masses.
But for simplicity of presentation, we will assume first/second generational
degeneracy. At first glance, one might expect that the generational
non-degeneracy would lead to violation of flavor-changing-neutral-current
(FCNC) bounds. The FCNC bounds mainly apply to first-second generation
nonuniversality\cite{Gabbiani:1996hi}.
However, the landscape itself allows a solution to the SUSY
flavor problem in that it statistically pulls all generations to large values
provided they do not contribute too much to $m_{weak}^{PU}$. This means the 3rd
generation is pulled to $\sim$ several TeV values whilst first and second
generation scalars are pulled to values in the $10-50$ TeV range.
The first and second generation scalar contributions to the weak scale
are suppressed by their small Yukawa couplings\cite{Baer:2012cf},
whilst their $D$-term contributions largely cancel under intra-generational universality\cite{Baer:2013jla}.
Their main influence on the weak scale then comes from two-loop RGE
contributions which, when large, suppress third generation soft term running
leading to tachyonic stop soft terms and possible charge-or-color breaking
(CCB) vacua which we anthropically veto\cite{Arkani-Hamed:1997opn,Baer:2000xa}.
These latter bounds are flavor independent so that first/second generation
soft terms are pulled to common upper bounds leading to a
quasi-degeneracy/decoupling solution to both the
SUSY flavor and CP problems\cite{Baer:2019zfl}.
Meanwhile, Higgs multiplets which live in different GUT representations
are expected to have independent soft masses $m_{H_u}$ and $m_{H_d}$\footnote{
  In models of local grand unification, the matter multiplets can live in the
  the $SO(10)$ spinor representations while the Higgs and gauge fields live in split multiplets due to their geography on the compactified manifold\cite{Nilles:2014owa}.}.
Thus, we expect a parameter space of the NUHM3 models as
\be
m_0(1)\simeq m_0(2),\ m_0(3),\ m_{H_u},\ m_{H_d},\ m_{1/2},\ A_0,\ \mu,\ {\rm and}\ m_A, 
\ee
where we used the EW minimization conditions allow for the more convenient
weak scale variables $\mu$ and $m_A$. For simplicity, sometimes we will
adopt $m_0(1)=m_0(2)=m_0(3)$ in which case we have the NUHM2 model.

Since $n=1$ is expected for KKLT moduli stabilization, then it is perhaps
more warranted to use the generalized mixed moduli anomaly (mirage) mediation
model GMM\cite{Baer:2016hfa}.\footnote{Also of interest is the more general anomaly-mediation model with separate bulk contributions to Higgs soft terms and a bulk $A$-term contribution\cite{Baer:2018hwa}. Then one can obtain naturalness within the AMSB framework;
  in this case, while winos are the lightest gauginos, the higgsinos are
  the lightest EWinos.} Applying landscape statistics to the GMM model, then
the magnitude of the mirage unification scale can be predicted. For results,
see Ref. \cite{Baer:2019tee}.

\begin{figure}[tbp]
\begin{center}
\includegraphics[height=0.35\textheight]{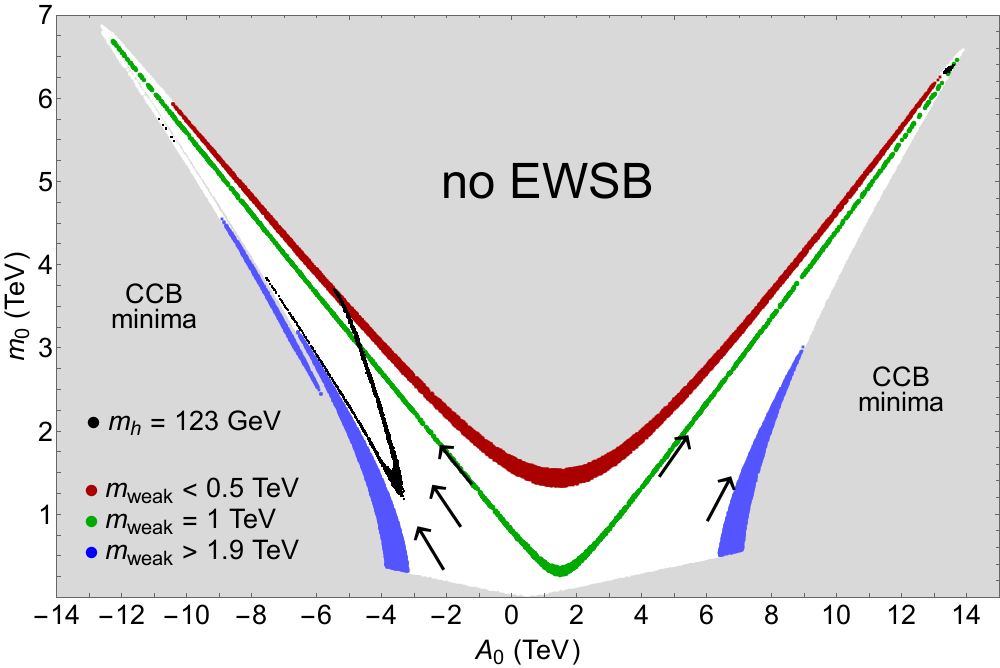}
\caption{Contours of $m_{weak}$ in the $A_0$ vs. $m_0$ plane
for $m_{1/2}=1$ TeV, $m_{H_u}=1.3 m_0$, $\tan\beta =10$ and $m_{H_d}=1$ TeV.
The arrows show the direction of statistical/anthropic pull on soft SUSY breaking terms.
Within the black contour is where $m_h>123$ GeV. 
There is also a slight black contour in the upper-right horn as well.
\label{fig:A0_m0}}
\end{center}
\end{figure}

In Fig. \ref{fig:A0_m0}\cite{Baer:2016lpj}, we show the 
$A_0$ vs. $m_0$ plane for the NUHM2 model with $m_{1/2}$ fixed at 1 TeV, $\tan\beta =10$
and $m_{H_d}=1$ TeV. We take $m_{H_u}=1.3 m_0$. 
The plane is qualitatively similar for different reasonable parameter choices. 
We expect $A_0$ and $m_0$ statistically to be drawn as large as possible
while also being anthropically drawn towards $m_{weak}\sim 100-200$ GeV, labelled as
the red region where $m_{weak}<500$ GeV. The blue region has $m_{weak}>1.9$ TeV and the green
contour labels $m_{weak}=1$ TeV. The arrows denote the combined 
statistical/anthropic pull on the soft terms: towards large soft terms but low $m_{weak}$.
The black contour denotes $m_h=123$ GeV with the regions to the upper left 
(or upper right, barely visible) containing larger values of $m_h$. 
We see that the combined pull on soft terms brings us to
the region where $m_h\sim 125$ GeV is generated. 
This region is characterized by highly mixed TeV-scale top
squarks\cite{Carena:2002es,Baer:2011ab}. 
If instead $A_0$ is pulled too large,
then the stop soft term $m_{U_3}^2$ is driven tachyonic resulting in charge and color
breaking minima in the scalar potential (labelled CCB). 
If $m_0$ is pulled too high for fixed $A_0$, then electroweak symmetry isn't even broken.

Next, we scan over parameter values
\bea
m_0(1,2)&:& 0-60\ {\rm TeV}\\
m_0(3)&:& 0.1-10\ {\rm TeV}\\
m_{H_u}&:& m_0(3)-2m_0(3)\\
m_{H_d}(\sim m_A) &:& 0.3-10\ {\rm TeV}\\
m_{1/2}&:& 0.5-3\ {\rm TeV}\\
-A_0&:& 0-50\ {\rm TeV}\\
\mu_{GUT}&:& \ fixed\\
\tan\beta &:& 3-60
\eea
The soft terms are all scanned according to  $f_{SUSY}\sim m_{soft}^{\pm 1}$
while $\mu$ is fixed at a natural value $\mu =150$ GeV.
For $\tan\beta$, we scan uniformly.
The goal is to take scan upper limits beyond those imposed by $f_{EWSB}$
so the plot upper bounds do not depend on scan limits.
The lower limits for the $n=-1$ case are selected in accord with
previous scans for $n=1$ with a draw to large soft terms just for
consistency. If we lower the lower bound scan limits,
then the $n=-1$ histograms will migrate to what becomes even worse
discord with experimental limits.

\begin{figure}[H]
\begin{center}
\includegraphics[height=0.22\textheight]{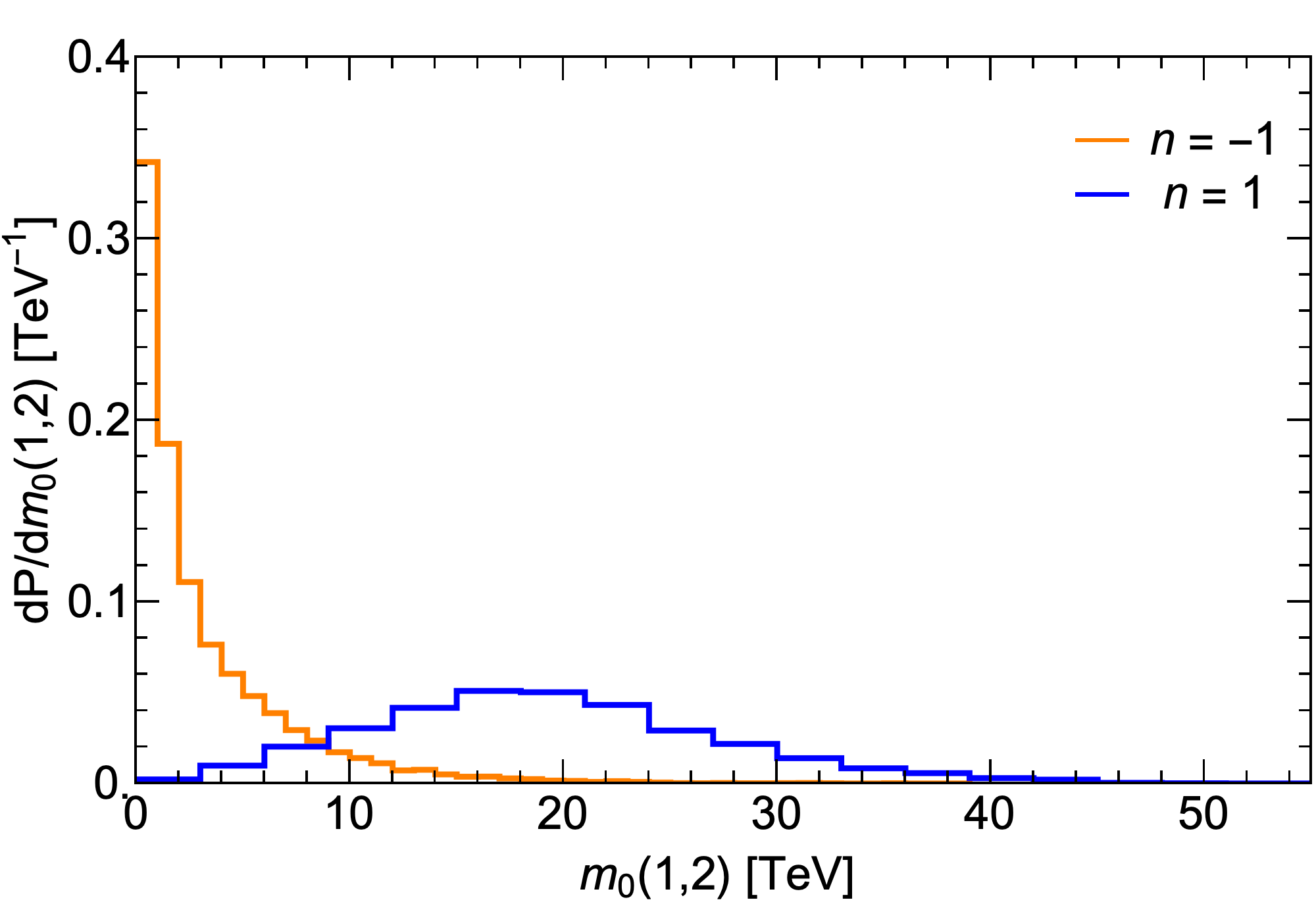}
\includegraphics[height=0.22\textheight]{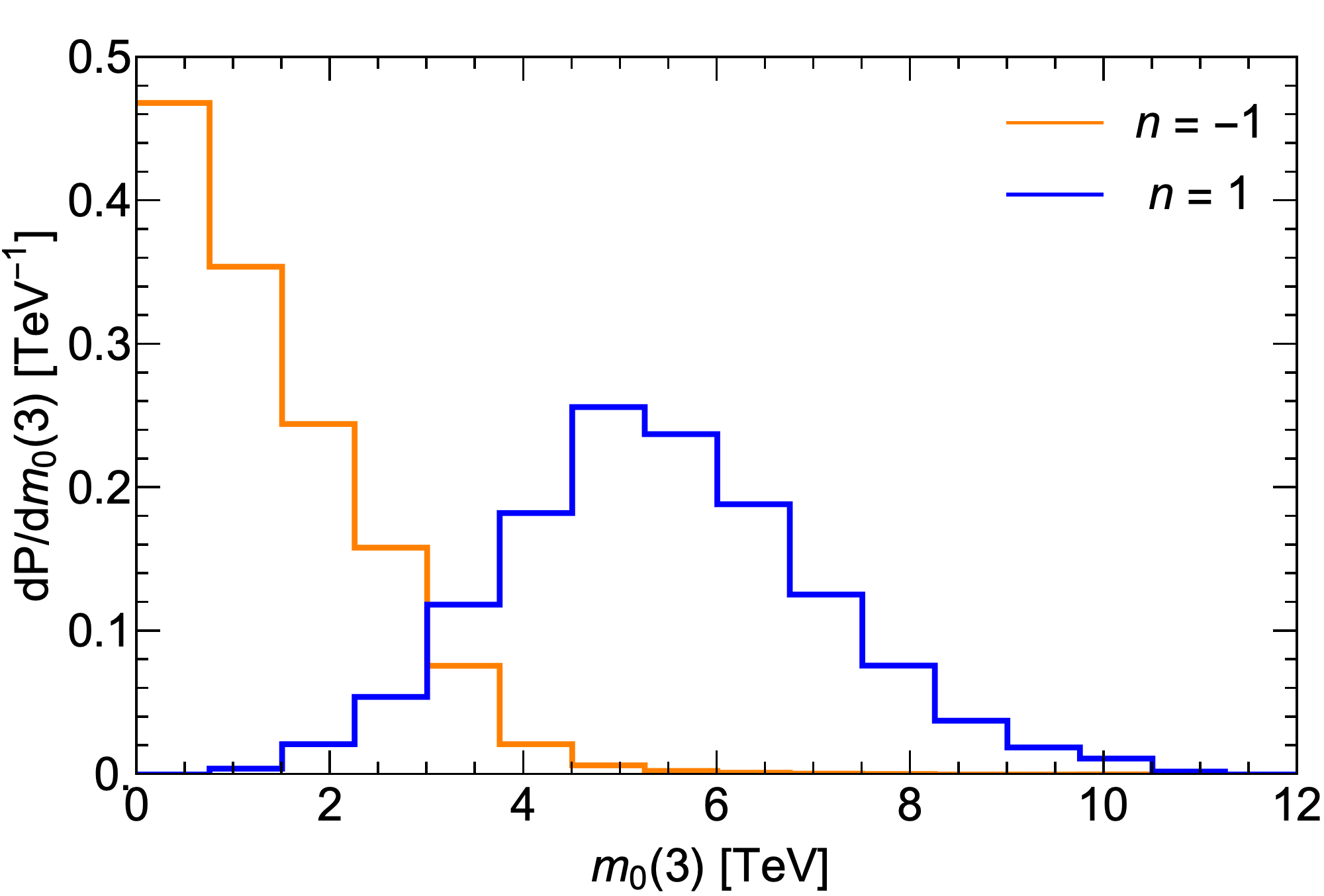}\\
\includegraphics[height=0.22\textheight]{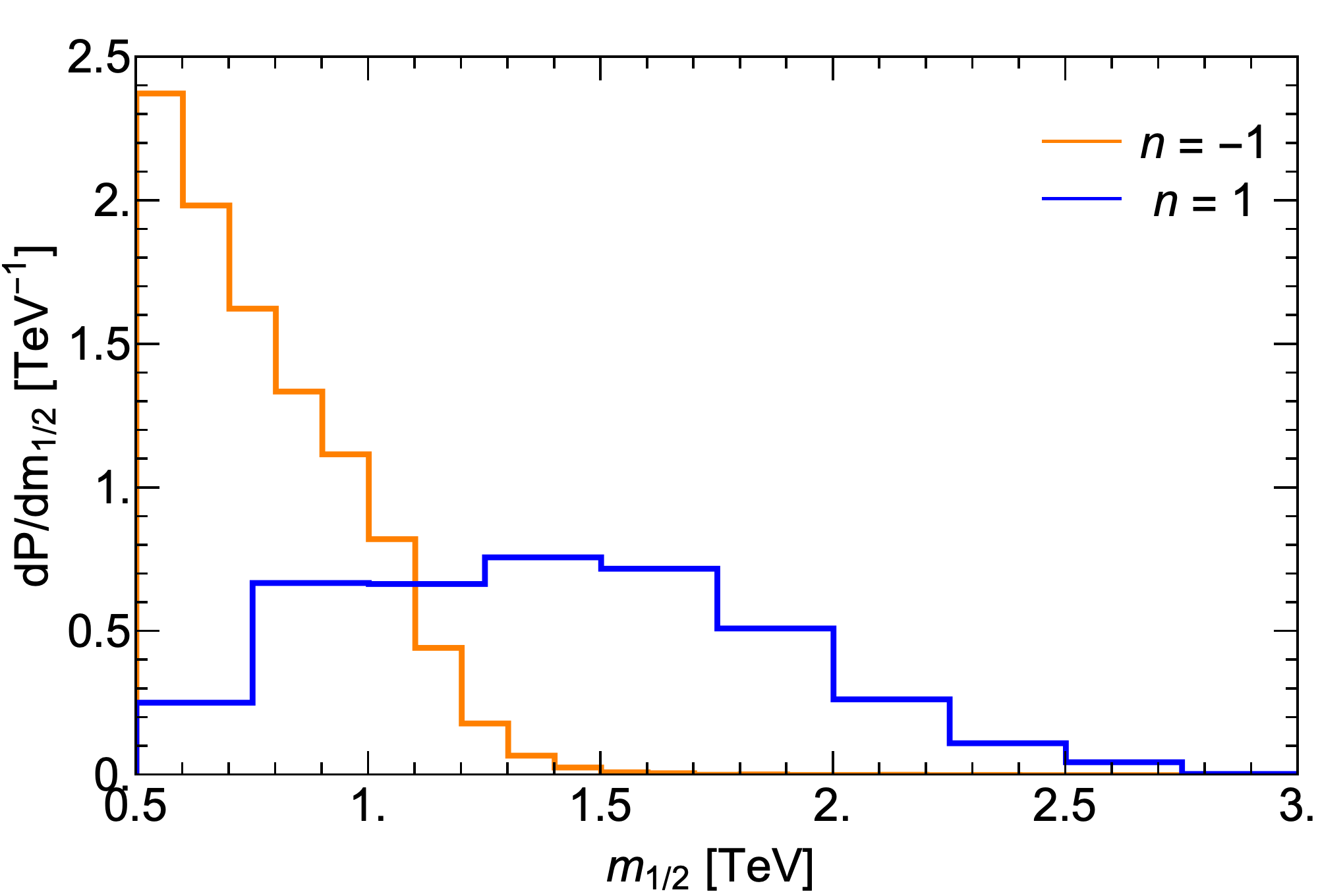}
\includegraphics[height=0.22\textheight]{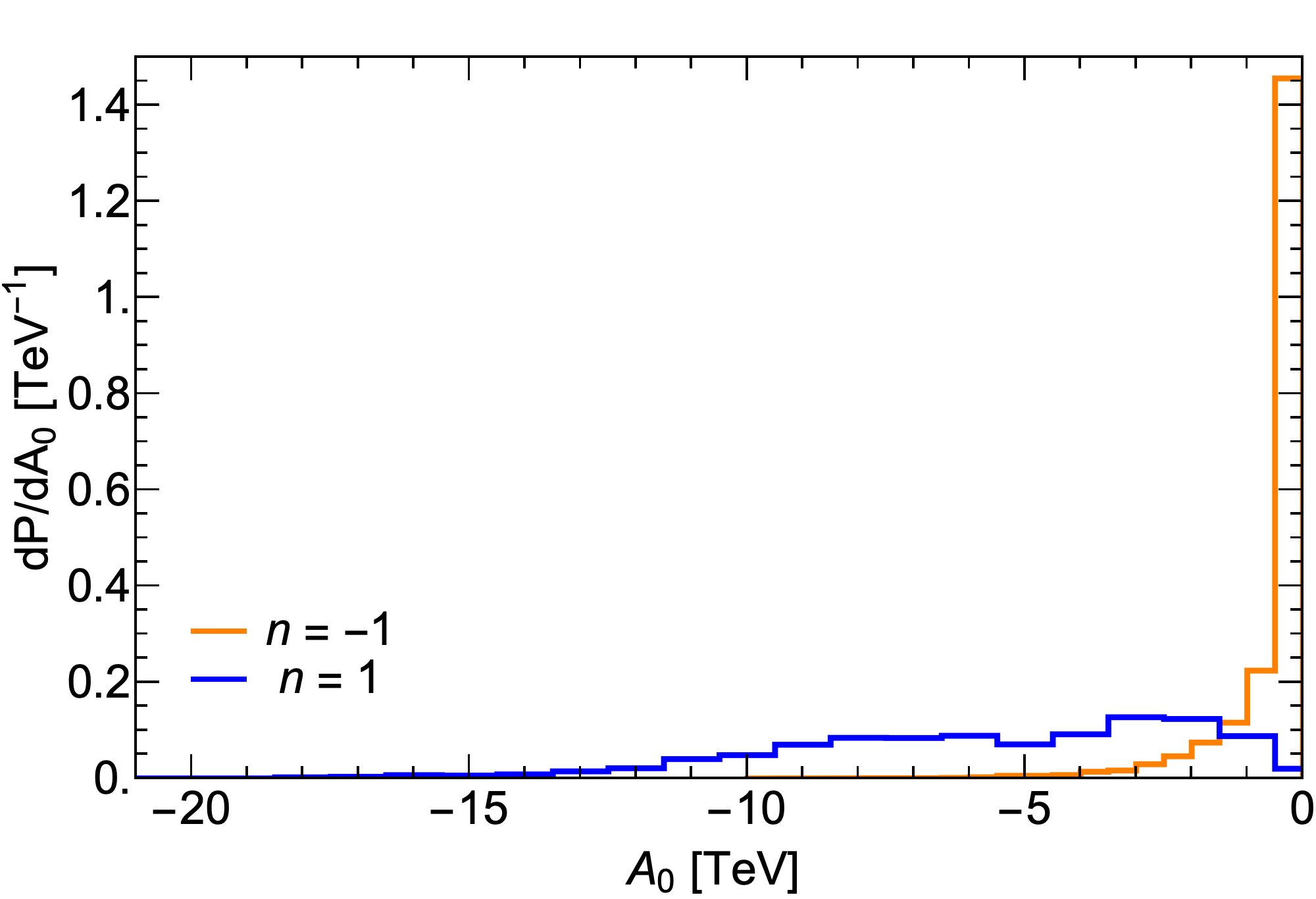}
\caption{Probability distributions for NUHM3 soft terms
{\it a}) $m_0(1,2)$, {\it b}) $m_0(3)$, {\it c}) $m_{1/2}$ and
{\it d}) $A_0$ from the $f_{SUSY}=m_{soft}^{\pm 1}$ distributions of soft terms
in the string landscape with $\mu =150$ GeV.
\label{fig:m0mhf}}
\end{center}
\end{figure}

In Fig. \ref{fig:m0mhf}, we show putative landscape distributions for
various NUHM3 parameters. In frame {\it a}), we show the distribution
for first/second generation scalar masses $m_0(1,2)$. For $n=1$, then we see
the probability distributions peaks around $m_0(1,2)\sim 20$ TeV but extends
as high as $\sim 45$ TeV. Such large first/second generation scalar masses
provide the decoupling/quasi-degeneracy soluton to the SUSY flavor and CP
problems. In contrast, for $n=-1$ then the distribution is sharply peaked
near 0 as expected. In frame {\it b}), we show the distribution in
third generation scalar soft mass $m_0(3)$.
Here, the $n=1$ distribution peaks at 5 TeV but runs as high as 10 TeV.
The $n=-1$ distribution again peaks at zero, which will lead to very light
third generation squarks. The distribution in $m_{1/2}$ shown in frame {\it c})
for $n=1$ peaks around $m_{1/2}\sim 1.5$ TeV leading to gaugino masses
typically beyond the present LHC limits. For $n=-1$, the distribution peaks at
low $m_{1/2}$ leading to gauginos that are typically excluded.
And in frame {\it d}) we see the distribution in trilinear soft term $-A_0$.
For $n=1$, the distribution has a double peak structure with most values
in the multi-TeV range leading to large stop mixing and consequently
cancellations in the $\Sigma_u^u(\tst_{1,2})$ and upift of $m_h$ to
$\sim 125$ GeV. For $n=-1$, then $A_0$ peaks around zero, and we expect
little stop mixing and lighter values of $m_h$.

\begin{figure}[H]
\begin{center}
\includegraphics[height=0.22\textheight]{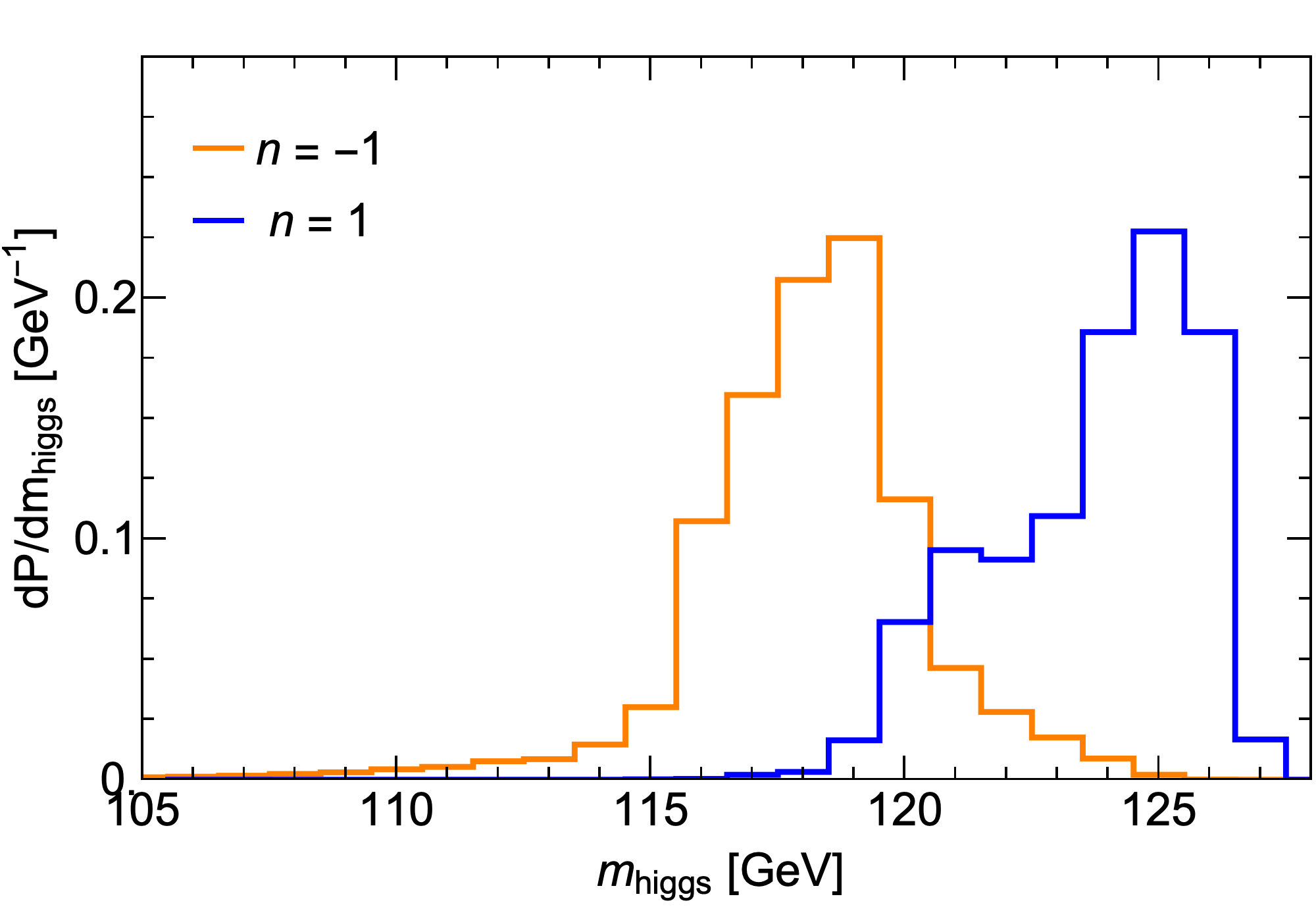}
\includegraphics[height=0.22\textheight]{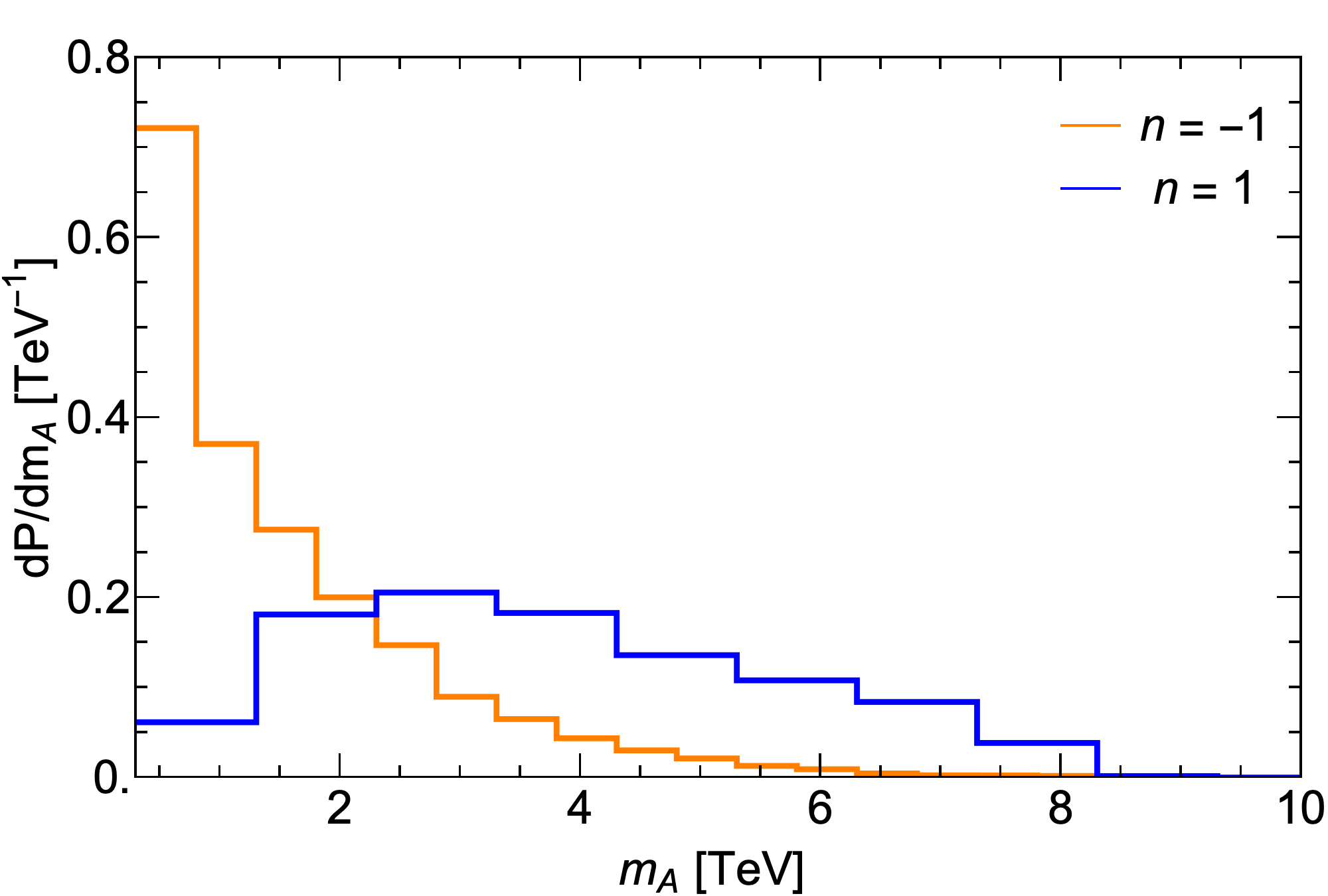}\\
\caption{Probability distributions for light
Higgs scalar mass
{\it a}) $m_h$ and pseudoscalar Higgs mass {\it b}) $m_A$
from the $f_{SUSY}=m_{soft}^{\pm 1}$ distributions of soft terms
in the string landscape with $\mu =150$ GeV.
\label{fig:higgs}}
\end{center}
\end{figure}

In Fig. \ref{fig:higgs}, we plot the landscape distributions for light and
heavy SUSY Higgs boson masses. In frame {\it a}), for $n=1$ we see a
distribution with a strong peak around $m_h\sim 124-126$ GeV in accord
with data. The distribution cuts off for $m_h\agt 127$ GeV because
otherwise the $\Sigma_u^u(\tst_{1,2})$ contributions become too large leading
to too large a value of $m_{weak}$ beyond the ABDS window. For $n=-1$, the
distribution peaks at $m_h\sim 118$ GeV with really no significant
probability beyond $m_h\sim 124$ GeV. This essentially rules out the $n=-1$
case. In frame {\it b}), the distribution in heavy pseudoscalar mass $m_A$
is shown. For $n=+1$, the distribution peaks at $m_A\sim 2.5$ TeV with
a distribution extending as high as $m_A\sim 8$ TeV. These values are well
beyond recent ATLAS search limits\cite{ATLAS:2020zms} from $H,\ A\to \tau\bar{\tau}$, which are plotted
in the $m_A$ vs. $\tan\beta$ plane. For $n=-1$, then we expect rather light
$m_A$, possibly at a few hundred GeV, leading to large light-heavy Higgs
mixing. This also seems in contradiction with LHC results which favor
a very SM-like light Higgs as expected in the decoupling limit.

\begin{figure}[H]
\begin{center}
\includegraphics[height=0.22\textheight]{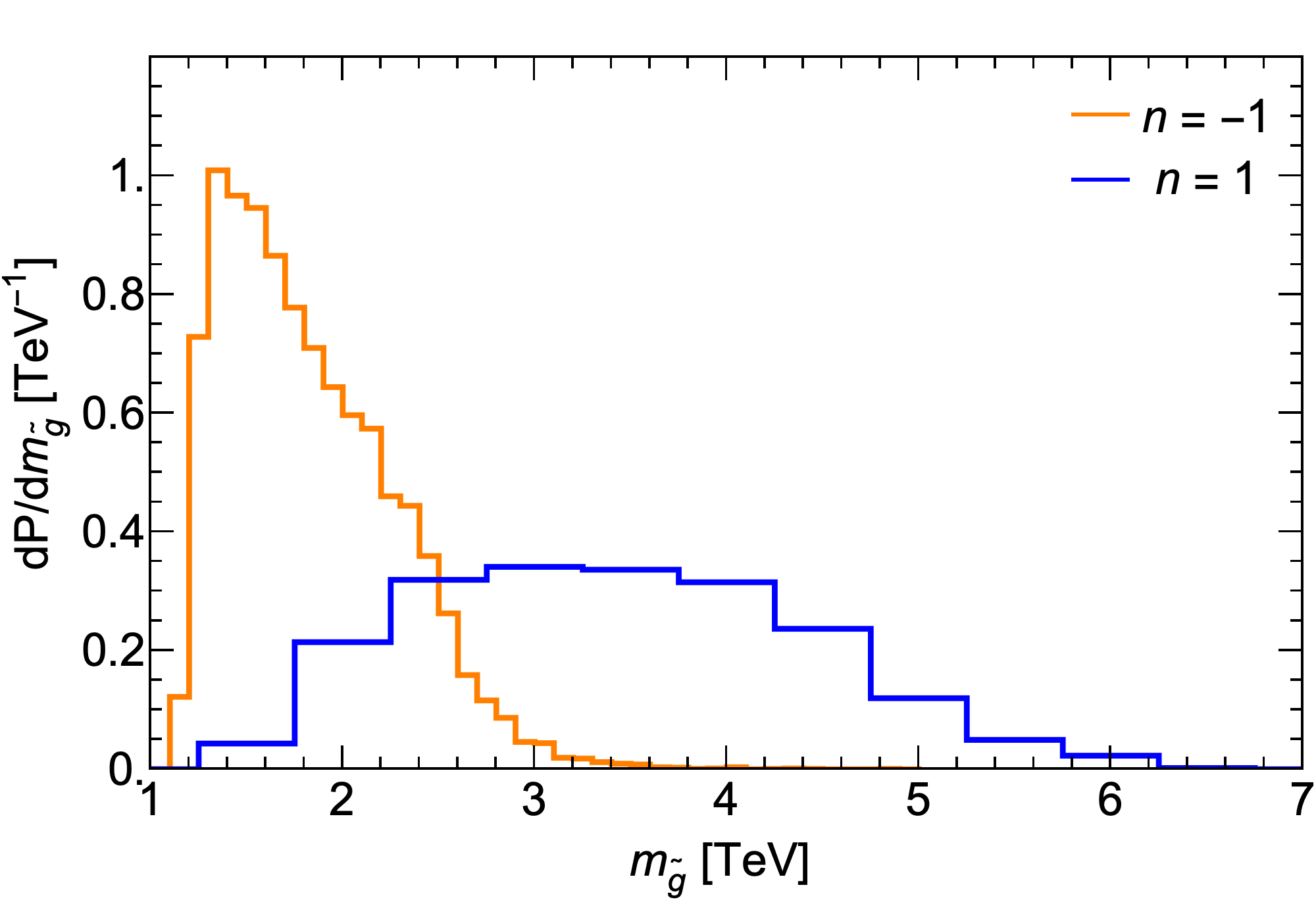}
\includegraphics[height=0.22\textheight]{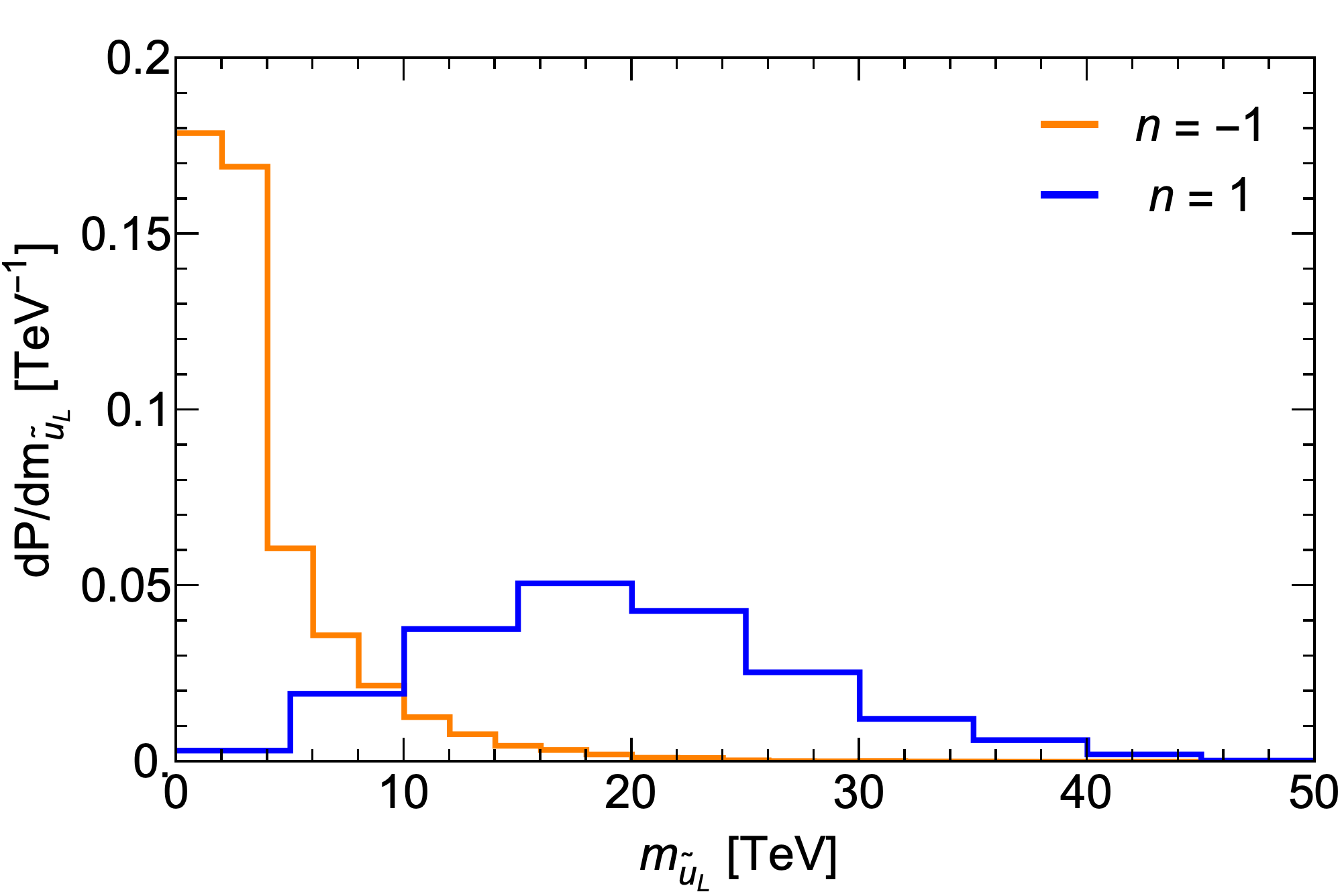}\\
\includegraphics[height=0.22\textheight]{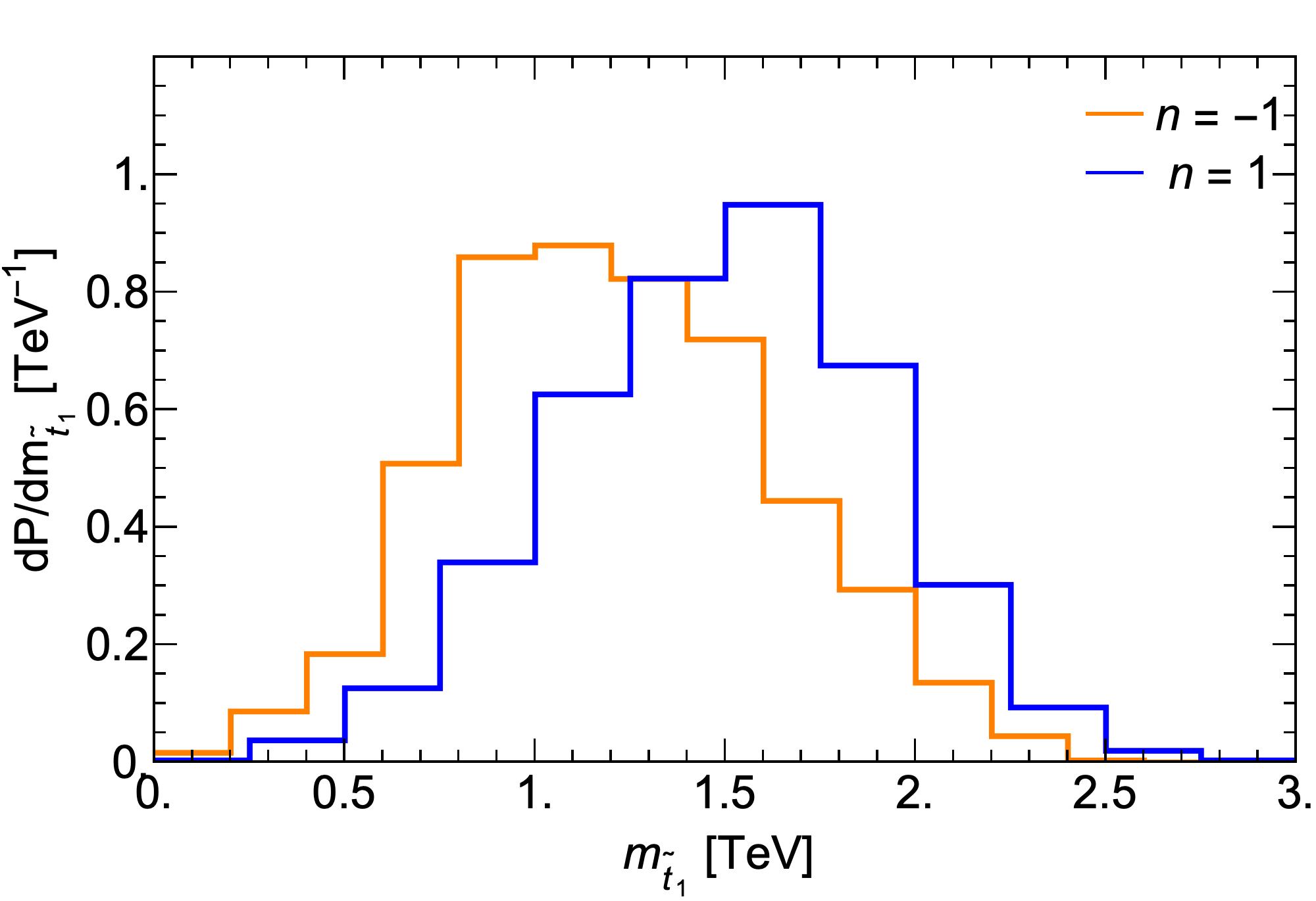}
\includegraphics[height=0.22\textheight]{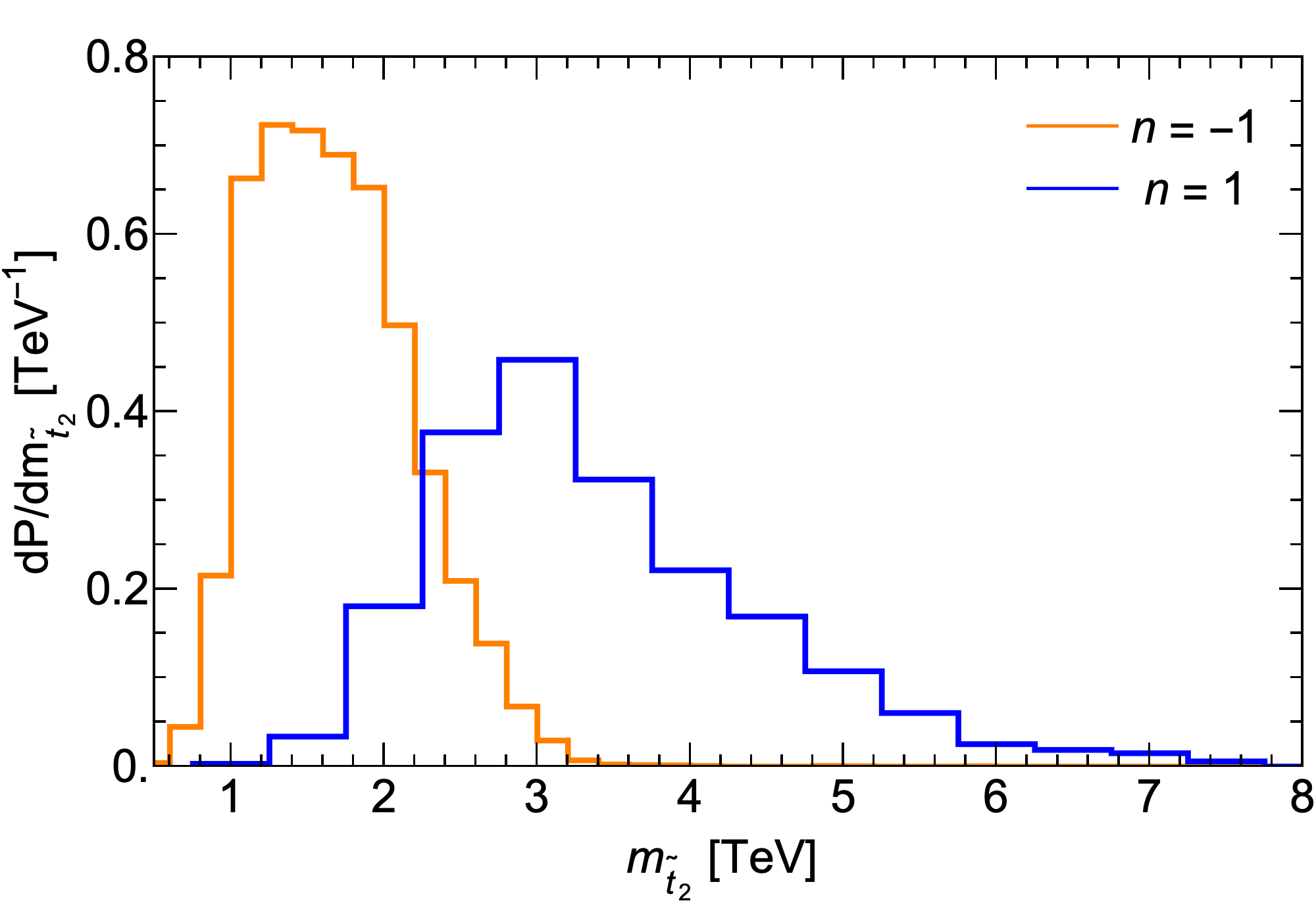}
\caption{Probability distributions for 
{\it a}) $m_{\tg}$, {\it b}) $m_{\tu_L}$, {\it c}) $m_{\tst_1}$ and
{\it d}) $m_{\tst_2}$ from the $f_{SUSY}=m_{soft}^{\pm 1}$ distributions of soft terms
in the string landscape with $\mu =150$ GeV.
\label{fig:mass}}
\end{center}
\end{figure}

In Fig. \ref{fig:mass}, we plot the expected strongly interacting
sparticle mass distributions from the landscape.
In frame {\it a}), we see for $n=1$ that $m_{\tg}$ peaks around
$m_{\tg}\sim 2.5-4$ TeV which is well beyond current LHC limits which
require $m_{\tg}\agt 2.1$ TeV. The upper distribution edge extends as far
as $m_{\tg}\sim 6$ TeV. In contrast, for the $n=-1$ distribution,
then the bulk of probability is below 2.1 TeV, although a tail does
extend somewhat above present LHC bounds. In frame {\it b}), we show the distribution in first generation squark mass $m_{\tu_L}$.
For $n=1$, the distribution peaks around $m_{\tq}\sim 20$ TeV but extends to
beyond 40 TeV. For $n=-1$, then squarks are typically expected at
$m_{\tq}\alt 1-2$ TeV and one would have expected squark discovery at LHC
(although a tail extends into the multi-TeV range).
In frame {\it c}), we show the light top squark mass distribution $m_{\tst_1}$.
Here, the $n=1$ distribution lies mainly between $1<m_{\tst_1}<\sim 2.5$ TeV
whereas LHC searches require $m_{\tst_1}\agt 1.1$ TeV. For $n=-1$, then
somewhat lighter stops are expected although there still is
about a 50\% probability to lie beyond LHC bounds on $m_{\tst_1}$.
In frame {\it d}), we show the distribution in $m_{\tst_2}$. For $n=1$,
we expect $m_{\tst_2}\sim 2-6$ TeV whilst for $n=-1$ then we expect instead
  that $m_{\tst_2}\sim 1-3$ TeV.

\section{Stringy naturalness}
\label{sec:stringy}

For the case of the string theory landscape, in Ref. \cite{Douglas:2004zg} 
Douglas has introduced the concept of {\it stringy naturalness}:
\begin{quotation}
{\bf Stringy naturalness:} the value of an observable ${\cal O}_2$ 
is more natural than a value ${\cal O}_1$ if more 
{\it phenomenologically viable} vacua lead to  ${\cal O}_2$ than to ${\cal O}_1$.
\end{quotation}

We can compare the usual naturalness measure $\Delta_{BG}$
to what is expected from stringy naturalness in the $m_0$ vs. $m_{1/2}$
plane\cite{Baer:2019cae}.
We generate SUSY soft parameters
in accord with Eq.~\ref{eq:dNvac} for values of $n=2n_F+n_D-1=1$ and 4.
The more stringy natural regions of parameter space are denoted by the higher
density of sampled points.
\begin{figure}[!htbp]
\begin{center}
\includegraphics[height=0.27\textheight]{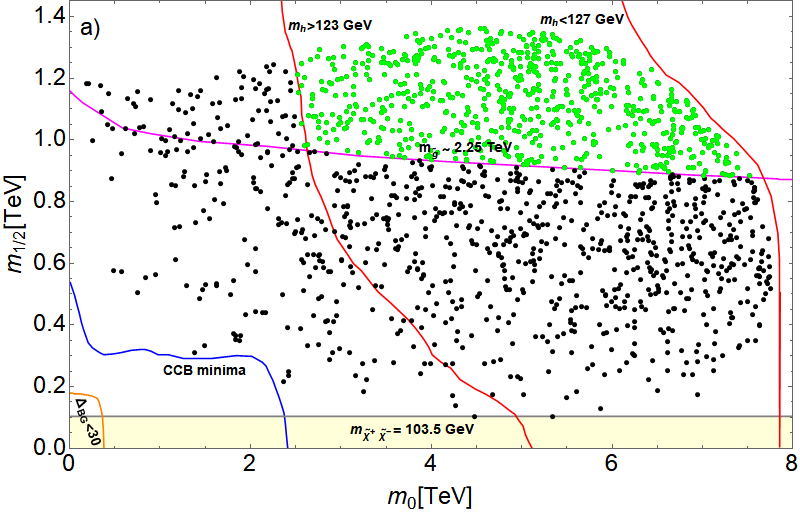}
\caption{The $m_0$ vs. $m_{1/2}$ plane of the NUHM2 model 
with $A_0=-1.6 m_0$, $\mu =200$ GeV and $m_A=2$ TeV and
an $n=1$ draw on soft terms,
The higher density of points denotes greater stringy naturalness.
The LHC Run 2 limit on $m_{\tg}>2.25$ TeV is shown by the magenta curve.
The lower yellow band is excluded by LEP2 chargino pair search limits.
The green points are LHC-allowed while black are LHC-excluded.
\label{fig:m0mhfn1}}
\end{center}
\end{figure}

In Fig. \ref{fig:m0mhfn1}, we show the stringy natural regions for the case
of $n=1$. 
Of course, no dots lie below the CCB boundary since such minima must be vetoed
as they likely lead to an unlivable pocket universe. 
Beyond the CCB contour, the solutions are in accord with livable vacua. 
But now the density of points {\it increases} with increasing 
$m_0$ and $m_{1/2}$ (linearly, for $n=1$), showing that the more stringy 
natural regions lie at the 
{\it highest} $m_0$ and $m_{1/2}$ values which are consistent with 
generating a weak scale within the ABDS bounds. 
Beyond these bounds, the density of points of course drops to zero 
since contributions to the weak scale exceed its measured value by
at least a factor of 4. 
There is some fluidity of this latter bound
so that values of $\Delta_{EW}\sim 20-40$ might also be entertained. 
The result that stringy naturalness for
$n\ge 1$ favors the largest soft terms (subject to $m_Z^{PU}$ not ranging too far from
our measured value) stands in stark contrast to conventional naturalness
which favors instead the lower values of soft terms. 
Needless to say, the stringy natural
favored region of parameter space is in close accord with LHC results in that
LHC find $m_h=125$ GeV with no sign yet of sparticles.

\begin{figure}[!htbp]
\begin{center}
\includegraphics[height=0.27\textheight]{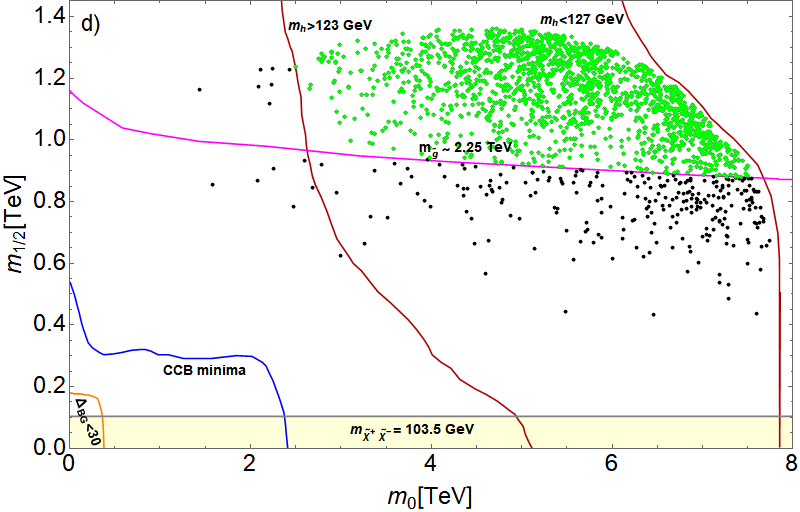}
\caption{The $m_0$ vs. $m_{1/2}$ plane of the NUHM2 model 
with $A_0=-1.6 m_0$, $\mu =200$ GeV and $m_A=2$ TeV and
an $n=4$ draw.
The higher density of points denotes greater stringy naturalness.
The LHC Run 2 limit on $m_{\tg}>2.25$ TeV is shown by the magenta curve.
The lower yellow band is excluded by LEP2 chargino pair search limits.
The green points are LHC-allowed while black points are LHC-excluded.
\label{fig:m0mhfn4}}
\end{center}
\end{figure}

In Fig. \ref{fig:m0mhfn4}, we show the same plane under an $n=4$ draw on 
soft terms. In this case, the density of dots is clearly highest 
(corresponding to most stringy natural) at the largest values of 
$m_0$ and $m_{1/2}$ as opposed to naive expectations where the most natural
regions are at low $m_0$ and $m_{1/2}$. 
In this sense, under stringy naturalness, a 3 TeV gluino is more natural 
than a 300 GeV gluino!

\section{Consequences of string landscape for SUSY collider searches}
\label{sec:colliders}

\begin{figure}[!htbp]
\begin{center}
\includegraphics[height=0.3\textheight]{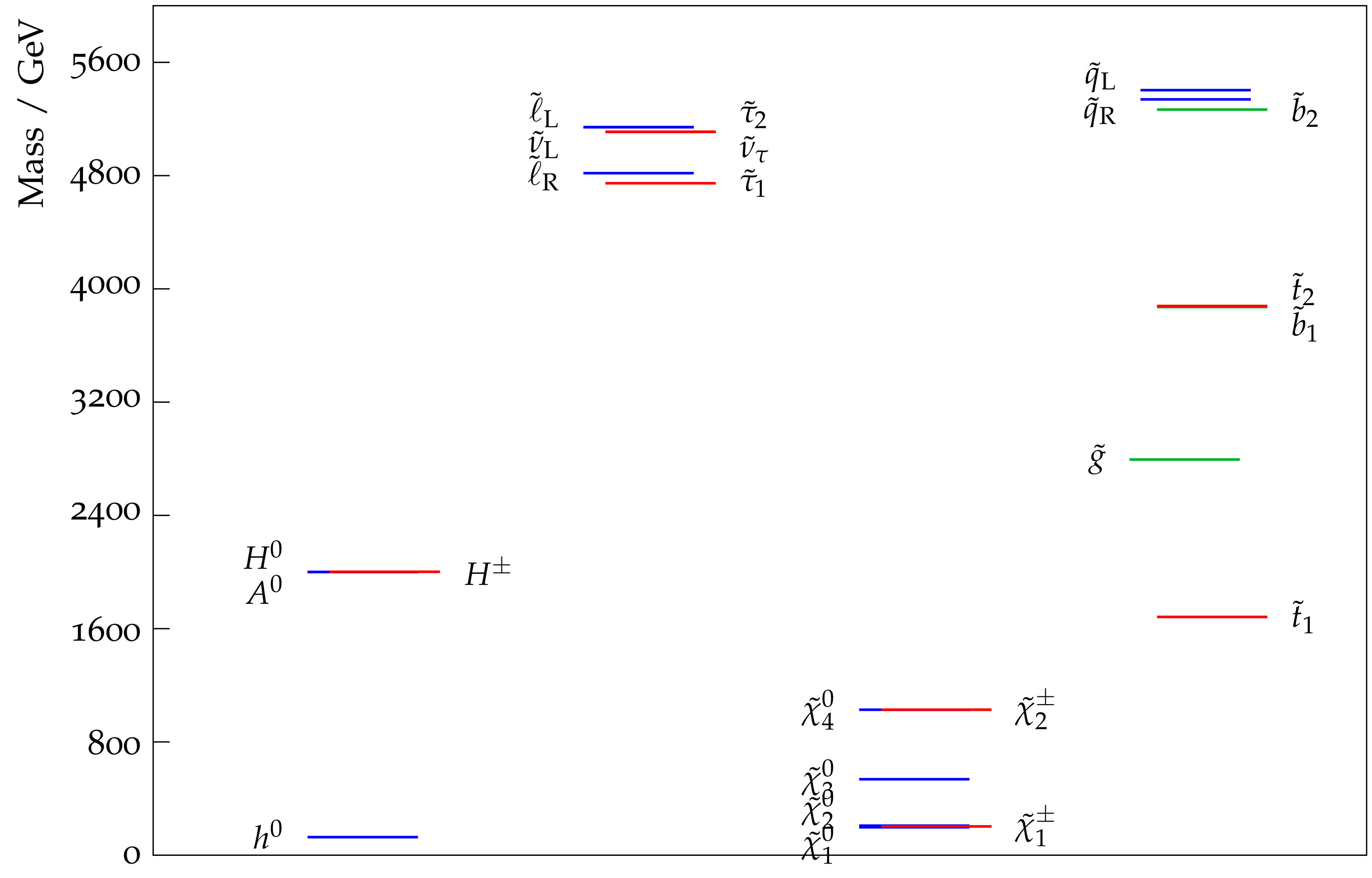}
\caption{Sparticle and Higgs mass spectra for a natural SUSY
benchmark point.
\label{fig:bm}}
\end{center}
\end{figure}

A figurative depiction of the expected sparticle and Higgs mass spectra
from the landscape is shown in Fig. \ref{fig:bm}.
Given such a spectra and the above distributions, we briefly describe expectations for SUSY at future collider options.
The big picture is that for a positive power-law draw to large soft terms from the string landscape, {\it then we expect a Higgs mass $m_h\sim 125$ GeV with
  sparticles beyond present LHC search limits}:
exactly what LHC is seeing so far
with $\sqrt{s}=13$ TeV and 139 fb$^{-1}$ of integrated luminosity.

\subsection{LHC}

\subsubsection{Light higgsinos}

Since the SUSY $\mu$ parameter is SUSY
conserving rather than SUSY breaking, it feeds mass to $W,\ Z$ and $h$
and also higgsinos (which mix with gauginos). Gaugino masses
are SUSY breaking and we expect the lightest EWinos to be mainly
higgsino-like, but {\it not} pure higgsino. We expect the higgsino-like
lightest EWinos $\tchi_1^\pm$ and $\tchi_{1,2}^0$ to have mass in the range
$\sim 100-350$ GeV.
Since the higgsino-like EWinos have very compressed spectra with mass gaps
$\sim 5-10$ GeV, then their visible decay products are expected to be very
soft and difficult to detect.
LHC searches for pair production of higgsino-like
EWinos was suggested in Ref. \cite{Baer:2011ec} as a probe of low $\mu$ via a
soft dimuon trigger. At present, the best search strategy seems to be to
look for $pp\to\tchi_2^0\tchi_1^0 +jet$ with $\tchi_2^0\to\ell^+\ell^-\tchi_1^0$\cite{Han:2014kaa,Baer:2014kya,Han:2015lma}. For Snowmass 2022,
the landscape parameter space has been mapped out in Ref. \cite{Baer:2020sgm}
and improved angular cuts for LHC searches have been proposed
in Ref. \cite{Baer:2021srt}.

\subsubsection{Gluino searches}

In the landscape, gluinos may range from $\sim 2-6$ TeV while top squarks
are in the $1-2.5$ TeV range. This means gluinos should decay to
top+stop or else three-body modes to top and bottom quarks\cite{Baer:1990sc,Baer:1998bj}.
The HL-LHC $5\sigma$ reach assuming 3000 fb$^{-1}$ is found to be
$m_{\tg}\sim 2.7$ TeV so there is some possibility these will
be discovered at LHC but more likely a higher energy hadron collider
with $\sqrt{s}\agt 30$ TeV will be needed\cite{Baer:2016wkz,Baer:2017pba,Baer:2018hpb}.

\subsubsection{Top squark pair searches}

In landscape SUSY, we expect light top squarks with mass
$m_{\tst_1}\sim 1-2.5$ TeV whilst the current LHC limits require
$m_{\tst_1}>1.1$ TeV. The reach of HL-LHC with 3000 fb$^{-1}$ extends to
$m_{\tst_1}\sim 1.3-1.7$ TeV. Thus, a higher energy hadron collider will be
needed to probe the entire expected light stop mass range\cite{Baer:2016wkz,Baer:2017pba,Baer:2018hpb}.
An important feature of landscape SUSY is that top squarks should be nearly
maximally mixed due to the required large weak scale value of
the trilinear soft term $A_t$.

\subsubsection{Same-sign diboson signature}

A qualitatively new signature for SUSY arises in natural models when $\mu$
is small. The reaction $pp\to\tchi_2^\pm\tchi_4^0$ whre the EWinos are
mainly wino-like can occur at high rates followed by decay to same-sign
$W$ bosons. This gives a unique SS dilepton plus MET signature
with minimal jet activity (just that from ISR) in distinction to
SS dileptons from gluino and squark production where substantial jet
activity is expected. Signal and background and LHC reach have been plotted out
in Ref's \cite{Baer:2013yha,Baer:2013xua,Baer:2017gzf}.

\subsection{Linear $e^+e^-$ collider}

\subsubsection{Direct production of light higgsinos}

Since light higgsinos are expeted in landscape SUSY with radiatively-driven
naturalness, then it makes sense to build something like the International
Linear $e^+e^-$ Collider (ILC). The ILC is touted as a Higgs factory, but if
$\sqrt{s}> 2m(higgsino)\sim 250-700$ GeV, then it may turn out
to be a higgsino factory as well\cite{Baer:2014yta}. The soft dileptons
arising from higgsino pair production ($e^+e^-\to \tchi_1^+\tchi_1^-$ and
$\tchi_1^0\tchi_2^0$) should be easily seen at ILC and their
invariant mass and energy spectra will allow precision determination of their
masses and mixings. One may even test gaugino mass unification\cite{Baer:2019gvu}.

\subsubsection{Precision Higgs measurements at a Higgs factory}

A primary goal of an $e^+e^-$ machine operating with $\sqrt{s}>m_Z+m_h$ is that
it can precisely measure Higgs boson properties, especially coupling strengths
$\kappa_i$ which could show deviations from SM predictions.
In landscape SUSY, since the soft terms are pulled to large values, one
gets decoupling and the expected Higgs couplings should look very
SM-like\cite{Bae:2015nva}.

\section{Consequences for WIMP and axion searches}
\label{sec:DM}

In natural SUSY with light higgsinos, the lightest neutralino is higgsino-like
and typically thermally underproduced by about a factor $5-15$.
If the underdensity of neutralinos is augmented by non-thermal higgsino production in the
early universe, then higgsino-only dark matter seems excluded by
direct and indirect DM detection experiments\cite{Baer:2018rhs}.
However, since axions are needed to solve the strong CP problem,
then a neutralino/axion dark matter mixture is to be expected\cite{Baer:2011hx}.
In  SUSY context with two Higgs doublets, the SUSY DFSZ axion model
is naturally expected\cite{Bae:2013bva,Bae:2013hma}, and axions tend to make up the bulk of the DM abundance.
The lower neutralino DM abundance allows light higgsinos to escape DD and IDD DM bounds\cite{Baer:2016ucr}.
The full mixed axion/higgsino DM
abundance reqires solution of eight coupled Boltzmann equations which include
the effects of axinos, axions, saxions and gravitinos\cite{Baer:2011uz,Bae:2014rfa}.
The axions are more difficult to detect than otherwise projected since now
higgsinos circulate in the axion-$\gamma$-$\gamma$ loop and reduce the
axion-photon coupling to tiny levels\cite{Bae:2017hlp}.

\section{Consequences for $(g-2)_\mu$}

The $a_\mu\equiv (g-2)_\mu/2$ anomaly has recently been reinforced by first data
from the Fermilab E989 experiment. To match the anomaly, SUSY theories
typically need light smuons and mu-sneutrinos.
The landscape tends to pull first/second generation sfermions into the 10-40 TeV
range so that $a_\mu$ should look very SM-like.
In this case, we would expect little or no anomaly\cite{Baer:2021aax}.

\section{Conclusions}
\label{sec:conclude}

The emergence of the string landscape of vacua has exciting consequences for
SUSY phenomenology (in addition to providing a solution to the CC problem).
With of order $10^{500}$ vacua to explore, statistical methods can be
brought to bear, and may even place string theory on a
long-awaited predictive footing. We present here a mini-review of our work
on the topic of stringy naturalness. We examined two main scenarios:
a power-law draw on soft terms to large ($n=1$) or small ($n=-1$) soft terms.
The former is motivated by the expectation of SUSY breaking by a single $F$ term
which is distributed uniformly as a complex number on the landscape,
and by KKLT moduli stabilization. The latter is motivated by an expectation
that the SUSY breaking scale is distributed uniformly over the decades of possibilities and arises in LVS moduli stabilization.
These statistical expectations must be tempered by the anthropic requirement
that the derived value for the weak scale in each pocket universe must lie within the ABDS window of values.
The LHC data clearly are in accord with the $n\ge 1$ statistics,
as they predict $m_h\sim 125$ GeV with sparticles typically
beyond present LHC search limits.
We also discussed these implications for LHC SUSY searches and for
WIMP and axion dark matter searches,
since we expect dark matter to consist of a WIMP/axion admixture.

{\it Acknowledgements:} 

This work is a Snowmass 2022 whitepaper.
This material is based upon work supported by the U.S. Department of Energy, 
Office of Science, Office of High Energy Physics under Award Number DE-SC-0009956 and DE-SC-001764.
Some of the computing for this project was performed at the OU
Supercomputing Center for Education and Research (OSCER) at
the University of Oklahoma (OU). The work of DS was supported by 
the Ministry of Science and Technology (MOST) of Taiwan under
Grant No. 110-2811-M-002-574.

%%%%%%%%%%%%%%%%%%%%%%%%%%%%%%%%%%%%%%%%%%%%%%%%%%%%%%

%\section*{References}
\bibliography{snow22}
\bibliographystyle{elsarticle-num}

\end{document}